\documentclass[fleqn,10pt]{wlscirep}
\usepackage{longtable}
\usepackage{bm}
\usepackage{dcolumn}
\usepackage{epsfig}
\usepackage{graphicx}
\usepackage{subfigure}
\usepackage{comment}
\usepackage{amsfonts}
\usepackage{amsmath}
\usepackage{amssymb}
\usepackage{subfigure}

\usepackage{color}

\newcommand{\be}{\begin{equation}}
\newcommand{\ee}{\end{equation}}
\newcommand{\bea}{\begin{eqnarray}}
\newcommand{\eea}{\end{eqnarray}}

\def\vec{\mathbf}

\usepackage{times}
\usepackage{appendix}

\begin{document}

\title{Strong magnetic frustration and anti-site disorder\\
       causing spin-glass behavior in honeycomb Li$_2$RhO$_3$}
\author[1,*]{Vamshi M.~Katukuri}

\author[1,+]{Satoshi Nishimoto}

\author[1]{Ioannis Rousochatzakis}

\author[2]{Hermann Stoll}

\author[1]{Jeroen van den Brink}

\author[1]{Liviu Hozoi}

\affil[1]{Institute for Theoretical Solid State Physics, IFW
Dresden, Helmholtzstr.~20, 01069 Dresden, Germany}
\affil[2]{Institute for Theoretical Chemistry, Universit\"{a}t Stuttgart, Pfaffenwaldring 55, 70550 Stuttgart, Germany}
\affil[*]{v.m.katukuri@ifw-dresden.de}
\affil[+]{s.nishimoto@ifw-dresden.de}

\begin{abstract}
With large spin-orbit coupling, the $t_{2g}^5$ electron configuration in $d$-metal oxides is prone to
highly anisotropic exchange interactions and exotic magnetic properties.
In $5d^5$ iridates, given the existing variety of crystal structures, the magnetic anisotropy can be
tuned from antisymmetric to symmetric Kitaev-type, with interaction strengths that outsize the
isotropic terms.
By many-body electronic-structure calculations we here address the nature of the magnetic exchange
and the intriguing spin-glass behavior of Li$_2$RhO$_3$, a $4d^5$ honeycomb oxide.
For pristine crystals without Rh-Li site inversion,
we predict a dimerized ground state as in the isostructural $5d^5$ iridate Li$_2$IrO$_3$,
with triplet spin dimers effectively placed on a frustrated triangular lattice.
With Rh-Li anti-site disorder, we explain the observed spin-glass phase as a superposition of different,
nearly degenerate symmetry-broken configurations.
\end{abstract}

\date\today
\maketitle

\section*{Introduction} 
For $d$-metal compounds with localized magnetic moments, basic guidelines to soothsay the sign
of the nearest-neighbor (NN) magnetic exchange interactions, i.~e., the Anderson-Goodenough-Kanamori
rules~\cite{Goodenough1958,Kanamori1959,anderson1959}, were laid down back in the 1950's.
With one single bridging anion and half-filled $d$ states these rules safely predict antiferromagnetic
(AF) exchange interactions, as is indeed encountered in numerous magnetic Mott insulators. 
It is however much harder to anticipate the sign of the couplings for geometries with two bridging
ligands and bond angles close to $90^{\circ}$.
Illustrative recent examples are the $5d$ honeycomb systems Na$_2$IrO$_3$ and Li$_2$IrO$_3$.
The signs of the NN Heisenberg $J$ and of the additional symmetric Kitaev anisotropy $K$ are
intensely debated in these iridates, with both $J\!<\!0$, $K\!>\!0$
\cite{Ir213_KH_chalopka_12,Ir213_andrade_14,Ir213_kee_14} and $J\!>\!0$, $K\!<\!0$
\cite{Ir213_jkj2j3_kimchi_2011,Ir213_jkj2j3_singh_2012,Ir213_choi_2012,Ir213_KH_mazin_2013,Ir213_rachel_14,Ir213_trebst_14}
sets of parameters being used to explain the available experimental data.

The sizable Kitaev interactions,
that is, uniaxial symmetric terms $KS_i^{z}S_{i+\mathbf{r}}^{z}$ ($\mathbf{r}\!=\!\frac{\mathbf{x}+\mathbf{y}}{\sqrt{2}}$)
that cyclically permute on the bonds of a particular hexagonal ring 
\cite{Kit_kitaev_06,Ir213_KH_jackeli_09,Ir213_KH_chaloupka_10}, are associated to strong frustration
effects and unconventional magnetic ground states displaying, for example, noncollinear order,
incommensurability, or spin-liquid behavior
\cite{Ir213_KH_chalopka_12,Ir213_andrade_14,Ir213_jkj2j3_kimchi_2011,Ir213_jkj2j3_singh_2012,Ir213_KH_chaloupka_10,Na2IrO3_vmk_14,Li2IrO3_vmk_14,Ir213_trebst_14,Ir213_kee_14,Ir213_rachel_14}.
Obviously, in the context of electronic-structure computational methods, such features cannot
be thoroughly addressed by periodic total-energy calculations for a given set of predetermined
spin configurations.
A much more effective strategy is to first focus on individual pairs of NN $d$-metal sites and 
obtain reliable values for the associated effective magnetic couplings by using {\it ab initio} 
many-body quantum chemistry (QC) machinery (for a recent review, see Ref.~\citeonline{qc_review_j_2014}).
The computed exchange parameters can be subsequently fed to effective spin Hamiltonians to be 
solved for larger sets of magnetically active lattice sites.
Such an approach, earlier, allowed us to establish the signs plus the relative strengths of the 
Heisenberg and Kitaev interactions in both Na$_2$IrO$_3$ and Li$_2$IrO$_3$ and to additionally
rationalize the qualitatively different types of AF orders in these two $5d^5$ honeycomb iridates
\cite{Na2IrO3_vmk_14,Li2IrO3_vmk_14}.

%
%
The related $4d^5$ honeycomb compound Li$_2$RhO$_3$ is even more puzzling because it features
no sign of long-range magnetic order.
Instead, an experimental study suggests the presence of a spin-glass ground state
\cite{Li2RhO3_HK_luo_13}.
While the spin-orbit couplings (SOC's) are still sizable for the $4d$ shell and may in principle
give rise on the honeycomb lattice to compelling Kitaev physics, to date no conclusive evidence
is in this respect available for Li$_2$RhO$_3$.
To shed light on the nature of the essential exchange interactions in Li$_2$RhO$_3$ we here
carry out detailed {\it ab initio} QC calculations.
We show that large trigonal splittings within the Rh $t_{2g}$ shell, comparable with the
strength of the SOC, dismiss a simple picture based on $j_{\rm eff}\!=\!1/2$ and
$j_{\rm eff}\!=\!3/2$ effective states \cite{SOC_d5_thornley68,book_abragam_bleaney,IrO_mott_kim_08,Ir213_KH_jackeli_09}.
The magnetic properties of the system can still be described, however, in terms of
${\tilde {S}}\!=\!1/2$ pseudospins.
The calculations earmark Li$_2$RhO$_3$ as a 4$d$-electron system with remarkably large
anisotropic magnetic couplings, in particular, FM Kitaev interactions of up to 10--15 meV.
The isotropic Heisenberg exchange, on the other hand, features opposite signs on the two sets
of structurally distinct links of Rh NN's.
This sign modulation of the NN Heisenberg interactions, with strong ferromagnetic (FM) $J$'s 
for one type of Rh-Rh bonds and weaker AF couplings for the other pairs of adjacent Rh sites,
enables the initial ${\tilde {S}}\!=\!1/2$ hexagonal network to be mapped
onto an effective model of spin-1 dimers on a frustrated triangular lattice. 
We further address the issue of Rh-Li anti-site disorder in samples of Li$_2$RhO$_3$.  
By exact-diagonalization (ED) calculations for an extended spin model that also includes second
and third neighbor couplings, we show that the experimentally observed spin-glass behavior
can be rationalized as a superposition of different nearly-degenerate symmetry-broken states
arising at finite concentration of in-plane spin vacancies.

\section*{Results}
\subsection*{Rh$^{4+}$ $4d^5$ electronic structure}
The tetravalent rhodium ions in Li$_2$RhO$_3$ display a 4$d^5$ valence electron configuration,
octahedral ligand coordination and bonding through two bridging ligands.
In the simplest picture, i.e., for sufficiently large Rh $t_{2g}$--$e_g$ splittings 
and degenerate $t_{2g}$ levels, the ground-state electron configuration at each 
site is a $t_{2g}^5$ effective $j_{\rm eff}\!=\!1/2$ spin-orbit doublet
\cite{SOC_d5_thornley68,book_abragam_bleaney,IrO_mott_kim_08,Ir213_KH_jackeli_09}.
For $5d^5$ ions in a variety of three-dimensional, layered, and chain-like oxides,
{\it ab initio} QC electronic-structure calculations yield excitation energies of 0.6--0.9
eV for the transitions between the $j\!\approx\!1/2$ and split $j\!\approx\!3/2$ levels
\cite{Ir213_rixs_gretarsson_2013,Ir3116_rixs_liu_2012,Ir227_hozoi_14,RhIr_vmk_IC_14} and
indicate values of 0.45--0.5 eV for the strength of the SOC $\lambda$, in agreement with
earlier estimates \cite{SOC_d5_andlauer76}.
Sharp features in the range of 0.6--0.9 eV are indeed found in the resonant x-ray scattering spectra
\cite{Ir214_rixs_jkim_12,Ir227_hozoi_14,Ir213_rixs_gretarsson_2013,Ir3116_rixs_liu_2012}.

The validity of the $j_{\rm eff}\!=\!1/2$ approximation for the ground state of Li$_2$RhO$_3$ is
however questionable since the SOC is substantially weaker for $4d$ elements.
Indeed our QC calculations (see Table~\ref{Rh213_dd}) indicate Rh $t_{2g}$ splittings $\delta\!\approx\!0.11$
eV, close to values of 0.14--0.16 eV estimated for $\lambda$ in various Rh$^{4+}$ oxides
\cite{Sr2RhO4_haverkort_2008,RhIr_vmk_IC_14}.
For the {\it ab initio} QC investigation we employed multiconfiguration complete-active-space
self-consistent-field (CASSCF) and multireference configuration-interaction (MRCI) calculations
\cite{book_QC_00}, 
see Supplemental Material (SM) and Refs.~\citeonline{RhIr_vmk_IC_14,Ir213_rixs_gretarsson_2013}.
With $\delta$ and $\lambda$ parameters of similar size, the $j_{\rm eff}\!=\!1/2$ and $j_{\rm eff}\!=\!3/2$
states are strongly admixed, as discussed in earlier work \cite{SOC_d5_thornley68,SOC_d5_hill71}
and illustrated in Table~\ref{Rh213_dd}.
In particular, for the relativistic ground-state wave function the $t_{2g}$ hole is not equally
distributed among the three Rh $t_{2g}$ levels as for the ``true'' $j_{\rm eff}\!=\!1/2$ ground
state \cite{SOC_d5_thornley68,book_abragam_bleaney} but displays predominant $d_{xy}$ character,
close to 60\%.

\subsection*{Magnetic couplings between two adjacent Rh$^{4+}$ ions}
Interestingly, while the results for the on-site $4d^5$ excitations are quite different as
compared to the $5d^5$ excitation energies \cite{Ir213_rixs_gretarsson_2013}, the computed
intersite effective interactions are qualitatively similar to those obtained for the $5d^5$
honeycomb iridate Li$_2$IrO$_3$ \cite{Li2IrO3_vmk_14}.
The intersite exchange couplings were estimated by MRCI+SOC calculations for embedded fragments
having two edge-sharing IrO$_6$ octahedra in the active region.
As described in earlier work \cite{Na2IrO3_vmk_14,Li2IrO3_vmk_14,Ba214_vmk_14} and in SM, the
{\it ab initio} QC data for the lowest four spin-orbit states describing the magnetic spectrum of
two NN octahedra is mapped in our scheme onto an effective spin Hamiltonian including both isotropic
Heisenberg exchange and symmetric anisotropy.
Yet the spin-orbit calculations, CASSCF or MRCI, incorporate all nine triplet and nine singlet
states that arise from the two-site $t_{2g}^5$--$t_{2g}^5$ configuration.

\begin{figure}[!ht]
\includegraphics[angle=0,width=0.72\columnwidth]{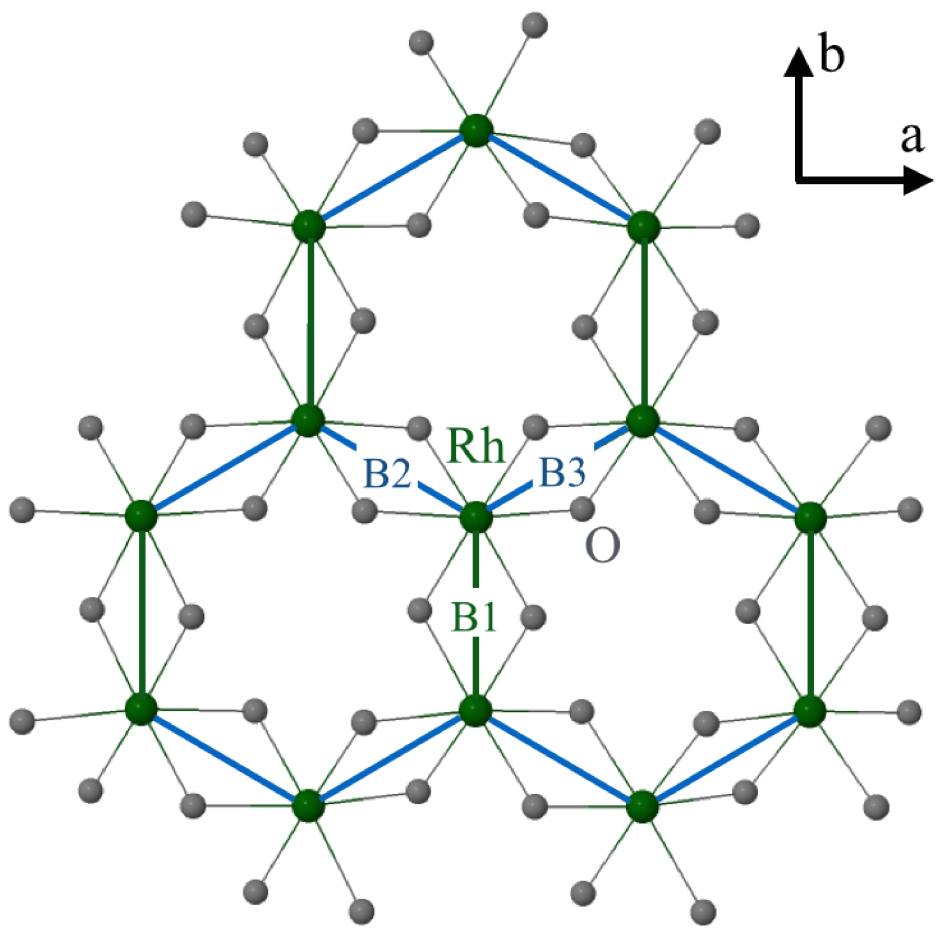}
\caption{
Layered network of edge-sharing RhO$_6$ octahedra in Li$_2$RhO$_3$.
The two distinct types, B1 and B2/B3, of NN two-octahedra units and the honeycomb lattice of Rh sites
are evidenced.
}
\label{213_structure}
\end{figure}

%
For on-site Kramers-doublet configurations, the most general symmetry-allowed form of the effective
spin Hamiltonian, for a pair of NN ions, is
\begin{equation}
{\mathcal H}_{ij}^{{C_{2h}}} = J\,{\tilde {\bf S}}_i\cdot {\tilde {\bf S}}_j
                             + K\,{\tilde S}_i^z {\tilde S}_j^z +
         \displaystyle\sum\limits_{\alpha<\beta}
         \Gamma_{\alpha\beta}({\tilde S}_i^{\alpha}{\tilde S}_j^{\beta} + {\tilde S}_i^{\beta}{\tilde S}_j^{\alpha})\,,
\label{eqn:Eqn1}
\end{equation}
where ${\tilde {\bf S}}_i$, ${\tilde {\bf S}}_j$ are 1/2 pseudospin operators, $J$ is the isotropic
Heisenberg interaction, $K$ the Kitaev coupling, and the $\Gamma_{\alpha\beta}$ coeeficients are offdiagonal
elements of the symmetric anisotropic exchange matrix with $\alpha,\beta\!\in\!\{x,y,z\}$.
The antisymmetric anisotropic term vanishes since the crystallographic data reported in 
Ref.~\citeonline{Li2RhO3_structure_2011} indicate overall $C_{2h}$ point-group symmetry for one block of NN
RhO$_6$ ocathedra, green (B1) bonds in Fig.\,\ref{213_structure}, and only slight deviations from
$C_{2h}$ for the other type of NN's, blue (B2 and B3) bonds in Fig.\,\ref{213_structure}.
For $C_{2h}$ symmetry of the Rh-Rh link, $\Gamma_{zx}\!=\!-\Gamma_{yz}$.
We note that in (\ref{eqn:Eqn1}) $\alpha$ and $\beta$ stand for components in the {\it local}, Kitaev
bond reference frame.
The $\vec{z}$ axis is here perpendicular to the Rh$_2$O$_2$ plaquette (see SI and
Refs.~\citeonline{Ir213_KH_jackeli_09,Na2IrO3_vmk_14,Li2IrO3_vmk_14}).


Relative energies for the four low-lying states describing the magnetic spectrum of two
NN octahedra and the resulting effective coupling constants are listed in Table~\ref{Rh213_mgn_coupl}.
%
%
For the effective picture of ${\tilde S}\!=\!1/2$ pseudospins assumed in Eq.\,(\ref{eqn:Eqn1}),
the set of four eigenfunctions contains the singlet
$\Phi^{ij}_{\mathrm{S}}\!=\!(\uparrow\downarrow\!-\!\downarrow\uparrow)/\sqrt2$ and the triplet components
$\Phi^{ij}_{\mathrm{1}}\!=\!(\uparrow\downarrow\!+\!\downarrow\uparrow)/\sqrt2$,
$\Phi^{ij}_{\mathrm{2}}\!=\!(\uparrow\uparrow\!+\!\downarrow\downarrow)/\sqrt2$,
$\Phi^{ij}_{\mathrm{3}}\!=\!(\uparrow\uparrow\!-\!\downarrow\downarrow)/\sqrt2$.
In $C_{2h}$ symmetry, the ``full" spin-orbit wave functions associated to $\Phi^{ij}_{\mathrm{S}}$,
$\Phi^{ij}_1$, $\Phi^{ij}_2$ and $\Phi^{ij}_3$ transform according to the $A_g$, $B_{u}$, $B_{u}$ and
$A_u$ irreducible representations, respectively.
Since two of the triplet terms may interact, the most compact way to express the eigenstates of
(\ref{eqn:Eqn1}) is then
$\Psi^{ij}_1\!=\! \Phi^{ij}_1\cos\alpha\!+\!i \Phi^{ij}_2\sin\alpha$,
$\Psi^{ij}_2\!=\!i\Phi^{ij}_1\sin\alpha\!+\!  \Phi^{ij}_2\cos\alpha$,
$\Psi^{ij}_3\!=\! \Phi^{ij}_3$ and
$\Psi^{ij}_{\mathrm{S}}\!=\!\Phi^{ij}_{\mathrm{S}}$.
The angle $\alpha$ parametrizes the amount of $\Phi^{ij}_1$--$\Phi^{ij}_2$ mixing, related to finite
off-diagonal $\Gamma_{\!\alpha\beta}$ couplings.
This degree of admixture is determined by analysis of the full QC spin-orbit wave functions.
The effective parameters provided in Table\,\ref{Rh213_mgn_coupl} are obtained for each type of Rh-Rh
link by using the $E_1$, $E_2$, $E_3$, $E_{\mathrm S}$ MRCI relative energies and the
$\Phi^{ij}_1$--$\Phi^{ij}_2$ mixing coefficients (see SM).
For B1 links, we find that both $J$ and $K$ are FM.
While by MRCI calculations $K$ always comes FM in spin-orbit coupled honeycomb systems
\cite{Na2IrO3_vmk_14,Li2IrO3_vmk_14}, the FM $J$ for the B1 bonds has much to do with the peculiar
kind of dependence on the amount of trigonal squashing of the oxygen octahedra and consequently
on the variation of the Rh-O-Rh angles of the Rh$_2$O$_2$ plaquette.
The latter increase to values larger than 90$^{\circ}$ for finite trigonal compression.
This dependence of the NN $J$ on the Rh-O-Rh bond angles is illustrated in Fig.\,2 for a simplified
structural model of Li$_2$RhO$_3$ where the Rh-O bond lengths are all the same, set to the average
bond length in the experimental crystal structure \cite{Li2RhO3_structure_2011}.
It is seen that $J$ displays a parabolic behavior, with a minimum of about --5 meV in the interval
92--93$^{\circ}$ and a change of sign to AF couplings around 96$^{\circ}$.
For the B1 Rh-Rh links, the Rh-O-Rh bond angle is 93.4$^{\circ}$, close to the value that defines the
minimum in Fig.\,2.
The difference between the $\approx$--5 meV minimum of Fig.\,2 and the $\approx$--10 meV result
listed in Table\,\ref{Rh213_mgn_coupl} comes from additional distortions of the O octahedra in the actual structure
(see the footnotes in Table\,\ref{Rh213_mgn_coupl} and Ref.\,\citeonline{Li2RhO3_structure_2011}), not included in the
idealized model considered for the plot in Fig.\,2.
An even stronger FM $J$ was computed for the B1 type bonds in the related compound Li$_2$IrO$_3$
\cite{Li2IrO3_vmk_14}.
In Na$_2$IrO$_3$, on the other hand, the Ir-O-Ir bond angles are $>$97$^{\circ}$ and the NN $J$
turns AF on all short Ir-Ir links \cite{Na2IrO3_vmk_14}.


For the B2 and B3 links, we derive a FM Kitaev term and an AF Heisenberg interaction, again
qualitatively similar to the QC data for Li$_2$IrO$_3$ \cite{Li2IrO3_vmk_14}.
We assign the AF value of the NN $J$ on the B2/B3 units to the slightly larger Rh-O-Rh bond angle,
which as shown in Fig.\,2 pulls the $J$ towards a positive value, and most importantly to additional
distortions that shift the bridging ligands on the Rh-O$_2$-Rh B2/B3 plaquettes in opposite senses
parallel to the Rh-Rh axis \cite{Li2RhO3_structure_2011}.
The role of these additional distortions on the B2/B3 units was analyzed in detail in
Ref.\,\citeonline{Na2IrO3_vmk_14} and shown to enhance as well the AF component to the intersite
exchange.

\subsection*{Effect of longer-range exchange interactions and occurence of spin-glass ground state}
For further insights into the magnetic properties of Li$_2$RhO$_3$, we carried out ED calculations for an 
extended spin Hamiltonian that in addition to the NN terms of Eq.(1) incorporates longer-range second- 
and third-neighbor Heisenberg interactions $J_2$ and $J_3$
\cite{Ir213_choi_2012,Ir213_jkj2j3_singh_2012,Ir213_jkj2j3_kimchi_2011,Ir213_andrade_14}.
We used clusters of 24 sites with periodic boundary conditions \cite{Ir213_KH_chalopka_12,Na2IrO3_vmk_14,Li2IrO3_vmk_14}
and the quantum chemically derived NN coupling constants listed in Table\,\ref{Rh213_mgn_coupl}. 
The static spin-structure factor 
$S({\bf Q})=\sum_{ij} \langle {\tilde {\bf S}}_i\cdot {\tilde {\bf S}}_j \rangle \exp[i {\bf Q}\cdot ({\bf r}_i-{\bf r}_j)]$ 
was calculated as function of variable $J_2$ and $J_3$ parameters.
For a given set of $J_2$ and $J_3$ values, the dominant order is determined according to the wave number 
{\bf Q} = ${\bf Q}_{max}$ providing a maximum value of $S({\bf Q})$.
The resulting phase diagram is shown in Fig.\,3(a).
Given the similar structure of the NN magnetic interactions, it is somewhat similar to that obtained
in our previous study on Li$_2$IrO$_3$ \cite{Li2IrO3_vmk_14}.
Six different regions can be identified for $|J_2|,|J_3|\!\lesssim\!6$ meV: FM, N\'eel, Kitaev spin
liquid, stripy, diagonal zigzag and incommensurate ${\bf Q}$ (ICx) phases. 
The Kitaev spin liquid, stripy and incommensurate phases in strongly spin-orbit coupled honeycomb $5d^5$ systems
were analyzed in a number of earlier 
studies \cite{Ir213_KH_chalopka_12,Ir213_jkj2j3_singh_2012,Ir213_jkj2j3_kimchi_2011,Ir213_kee_14,Na2IrO3_vmk_14}.
The detailed nature of the diagonal zigzag and incommensurate ICx ground states for large FM $J$ on
one set of NN links was described in Ref.\,\citeonline{Li2IrO3_vmk_14}.
Remarkably, for $J({\mathrm{B1}})$ much larger than $K({\mathrm{B1}})$ and $J({\mathrm{B2}})$, the initial 
hexagonal $\tilde S\!=\!1/2$ lattice can be mapped onto an effective triangular model of {\it triplet} dimers
on the B1 bonds \cite{Li2IrO3_vmk_14}.


Since $J_2$ and $J_3$ are expected to be AF in honeycomb $d^5$ oxides \cite{Ir213_choi_2012,Ir213_jkj2j3_singh_2012},
the most likely candidate for the magnetic ground state of ``clean'' crystals of Li$_2$RhO$_3$,
according to our results, is the diagonal zigzag state (see Fig.\,3) and is found to be stable in a wide
region of $J_2\!\gtrsim\!0$ and $J_3\!\gtrsim\!0$.
Experimentally, however, a spin-glass ground state was determined, with a spin freezing temperature of
$\sim\!6$\,K \cite{Li2RhO3_HK_luo_13}.
As possible cause of the observed spin-glass behavior in Li$_2$RhO$_3$ we here investigate the role of
Li-Rh site intercalation.
Significant disorder on the cation sublattice is a well known feature in Li$_2${\it M}O$_3$ compounds.
A typical value for the degree of Li$^+$--{\it M}$^{4+}$ site inversion in these materials is 10--15$\%$
\cite{kobayashi_95,Ir213_ryoji_2003}.
Partial substitution of the ``in-plane'' Rh$^{4+}$ ions by nonmagnetic Li$^+$ species introduces spin
deffects in the $\tilde S\!=\!1/2$ honeycomb layer.
On the 24-site cluster employed for our ED calculations, 10--15$\%$ site inversion translates in
replacing two $\tilde S\!=\!1/2$ centers by vacancies.
Hereafter, we denote the two spin defects as $p_1$ and $p_2$.

The effect of spin vacancies on the static spin-structure factor in the diagonal zigzag phase
($J_2\!=\!J_3\!=\!3$) is shown in Fig.\,3(d),(e). 
For comparison, the static spin-structure factor is also plotted in Fig.\,3(c) for the ideal case
without spin defects.
In the absence of ``defects'', the ground state is characterized in the bulk limit by symmetry-broken
long-range order with either
${\bf Q}_1\!=\!(\pm\pi,\pm\frac{\pi}{\sqrt{3}})$ or
${\bf Q}_2\!=\!(\pm\pi,\mp\frac{\pi}{\sqrt{3}})$ wave vectors.
Since the two symmetry-breaking states are degenerate, the structure factor displays four peaks, at
${\bf Q}\!=\!{\bf Q}_1$ and ${\bf Q}\!=\!{\bf Q}_2$ [see Fig.\,3(c)].
However, by introducing spin vacancies, the degeneracy may be lifted via impurity pinning effects.
For example, when the two defects occupy positions 17 and 20 [$(p_1,p_2)\!=\!(17,20)$, see Fig.\,3(b)]
the spin structure defines one of the symmetry-breaking states with ${\bf Q}\!=\!{\bf Q}_1$ [Fig.\,3(d)];
likewise, defects at $(p_1,p_2)\!=\!(17,18)$ yield a state with ${\bf Q}\!=\!{\bf Q}_2$  [Fig.\,3(e)].
In other words, two different kinds of dominant short-range order can be obtained with anti-site disorder.
The ``locally'' favored symmetry-breaking direction depends on the relative positions of the spin
vacancies.
In a macroscopic system, such ``local'' domains displaying different symmetry-breaking ordering directions
are randomly distributed.
Additional {\it frustration} is expected to arise because it is not possible to match two differently
ordered domains without an emerging ``string''.
It is therefore likely that by creating some amount of spin defects the long-range zigzag order disappears
and the resulting state is perceived as a spin glass at low temperature.
A similar mechanism was proposed for the isotropic Heisenberg-Kitaev and $J_1$-$J_2$-$J_3$ models
\cite{Ir213_andrade_14}.

An early well known example of frustration induced through the competition between two different,
degenerate spin configurations is the two-dimensional Ising model on a square lattice with randomly 
distributed, competing FM and AF bonds~\cite{Vannimenus_77}. 
To investigate how the diagonal zigzag state is destroyed by increasing the concentration of spin
defects, we also studied a simplified Ising model with $J\!=\!-\infty$ for the B1 bonds, $J_2\!=\!J_3$
and all other interactions set to zero.
This is a reasonable approximation for the honeycomb layers of Li$_2$RhO$_3$ since 
the diagonal zigzag phase essentially consists of alternating spin-up and spin-down chains [see
sketch in Fig.\,3(a)].
Spin structures obtained this way for various spin-defect concentrations $x$ are shown in Fig.\,3(f)-(i).
For $x\!=\!0$ the symmetry-breaking diagonal zigzag state is realized, with degenerate ${\bf Q}\!=\!{\bf Q}_1$
and ${\bf Q}\!=\!{\bf Q}_2$ spin structures.
At finite, low concentration $x\!\sim\!2\%$ those two configurations are no longer degenerate since
one of them features slightly lower ground-state energy.
We still have in this case a ``macroscopically stable'' ground state.
At intermediate defect concentration $x\!\sim\!7\%$ the long-range order is in a strict sense destroyed.
However, the large domain walls with either ${\bf Q}\!=\!{\bf Q}_1$ or ${\bf Q}\!=\!{\bf Q}_2$ seem
to survive. 
At higher concentration $x\!\sim\!12\%$ the long-range order disappears completely.
Moreover, we can now identify a mixture of local structures with different symmetry-breaking directions
[see Fig.\,3(i)].

\section*{Discussion}
In sum, we have calculated the microscopic neareast-neighbor magnetic interactions between effective
1/2 spins in Li$_2$RhO$_3$ and uncovered a substantial difference between the two types of bonds
that are present: one is dominated by Heisenberg and the other by Kitaev types of couplings.
The latter give rise to strong frustration, even if the interactions are predominantly
ferromagnetic.
In this setting we have additionally considered the effect of the presence of anti-site disorder.
Experimentally the in-plane spin-defect concentration in Li$_2$RhO$_3$ has been estimated as
$x\!=$10--15$\%$ \cite{kobayashi_95,Ir213_ryoji_2003}.
Based on our theoretical findings it is likely that the observed spin-glass behavior arises
from the combination of such anti-site disorder and strongly frustrating magnetic interactions,
in particular, the different Kitaev/Heisenberg dominated magnetic bonds and the Ising-like
physics associated with the triplet dimer formation that results from there.

Our combined {\it ab initio} and effective-model calculations on both Li$_2$RhO$_3$ and related
$d^5$ honeycomb iridates \cite{Ir213_rixs_gretarsson_2013,Na2IrO3_vmk_14,Li2IrO3_vmk_14} indicate
that a description in terms of on-site 1/2 pseudospins can well account for the diverse
magnetic properties of these systems.
While alternative models rely on the formation of delocalized, quasimolecular orbitals 
\cite{Mazin_PRL_Na213_2012,213_hexagMO_2013} and for Li$_2$RhO$_3$ downplay the role of spin-orbit
interactions \cite{213_hexagMO_2013}, here we show that the latter give rise in Li$_2$RhO$_3$
to anisotropic Kitaev interactions the same magnitude as in $5d$ iridates
\cite{Na2IrO3_vmk_14,Li2IrO3_vmk_14,Ir213_KH_gretarsson_2013}.
That happens in spite of having a Rh $t_{2g}$ splitting $\delta$ and a spin-orbit coupling $\lambda$
of similar magnitude, the same way similar sets of Ir $\delta$ and $\lambda$ parameters 
in CaIrO$_3$ \cite{Ir113_bogdanov_12} still generate symmetric anisotropic exchange terms in the
range of 10 meV (work is in progress). 


\section*{Methods}
The {\sc molpro} QC package was employed for all {\it ab initio} calculations~\cite{MOLPRO_WIREs}. 
To analyze the electronic ground state and the nature of the $d$-$d$ excitations, 
a cluster consisting of one reference
RhO$_6$ octahedron plus three NN RhO$_6$ octahedra and 15 nearby Li ions was used.
The magnetic spectrum for two Rh$^{4+}$ ions was obtained from calculations on a cluster containing two 
reference 
and four NN RhO$_6$ octahedra plus the surrounding 22 Li ions, see SI for details.
The farther solid-state environment was in both cases modeled as a finite array of point charges fitted 
to reproduce the crystal Madelung field in the cluster region. 
The spin-orbit treatment was carried out according to the procedure described in
Ref.~\citeonline{SOC_molpro}, using spin-orbit pseudopotentials for Ir. 

\section*{Acknowledgements}
We thank V.~Yushankhai, Y.~Singh, N.~A.~Bogdanov, and U.~K.~R\"{o}{\ss}ler for useful discussions.
L.~H. acknowledges financial support from the German Research Foundation (Deutsche
Forschungsgemeinschaft, DFG).

\section*{Author contributions}
V.M.K. carried out the {\it ab initio} calculations and subsequent mapping of the {\it ab initio} results onto the effective spin 
Hamiltonian, with assistance from L.H., H.S. and I.R.
S.N. performed the exact-diagonalization calculations. 
V.M.K., S.N., J.v.d.B. and L.H. analyzed the data and wrote the paper, with contributions from 
all other coauthors.

\section*{Additional information}
\textbf{Competing financial interests:} Authors have no competing financial interests.


\begin{table}[!ht]
\caption{
Rh$^{4+}$ $t_{2g}^5$ states in Li$_2$RhO$_3$, with composition of the wave functions (hole picture)
and relative energies (meV).
CASSCF results without and with SOC are shown.
Only the three Rh $t_{2g}$ orbitals were active \cite{book_QC_00} in CASSCF.
By subsequent MRCI calculations, the relative energies of these states change to 0, 85, 95 
without SOC and 0, 235, 285 meV with SOC included.
Only one component of the Kramers' doublet is shown for each spin-orbit wave function.
}
\label{Rh213_dd}
\begin{tabular}{llr}
\hline
\hline\\[-0.25cm]
$t_{2g}^5$ states  &Relative   &Wave-function composition\\
                   &energies   &(normalized weights)\\
\hline\\[-0.25cm]
CASSCF\,:\\
$|\phi_1\rangle$  &0           &$0.84\,|xy\rangle + 0.08\,|yz\rangle + 0.08\,|zx\rangle$\\[0.1cm]
$|\phi_2\rangle$  &107         &                   $0.50\,|yz\rangle + 0.50\,|zx\rangle$\\[0.1cm]
$|\phi_3\rangle$  &110         &$0.16\,|xy\rangle + 0.42\,|yz\rangle + 0.42\,|zx\rangle$\\[0.20cm]

CASSCF+SOC\,:\\
$|\psi_1\rangle$  &0                 &$   0.44\,|\phi_1,\uparrow\rangle + 0.15\,|\phi_1,\downarrow\rangle$ \\
&				     &$ + 0.09\,|\phi_2,\uparrow\rangle + 0.12\,|\phi_2,\downarrow\rangle$ \\
&				     &$ + 0.05\,|\phi_3,\uparrow\rangle + 0.15\,|\phi_3,\downarrow\rangle$\\[0.1cm]
$|\psi_2\rangle$  &210               &$   0.03\,|\phi_1,\uparrow\rangle + 0.38\,|\phi_1,\downarrow\rangle$\\ 
& 				     &$ + 0.29\,|\phi_2,\uparrow\rangle$                                  \\
&				     &$ + 0.27\,|\phi_3,\uparrow\rangle + 0.03\,|\phi_3,\downarrow\rangle$\\[0.1cm]
$|\psi_3\rangle$  &265               &$   0.17\,|\phi_2,\uparrow\rangle + 0.32\,|\phi_2,\downarrow\rangle$\\
&			             &$ + 0.08\,|\phi_3,\uparrow\rangle + 0.43\,|\phi_3,\downarrow\rangle$ \\
%
%
%
%
\hline
\hline
\end{tabular}
\end{table}

\begin{table}[!ht]
\caption{
Relative energies of the four low-lying magnetic states and the associated effective exchange
couplings (meV) for two NN RhO$_6$ octahedra in Li$_2$RhO$_3$.
Two distinct types of such [Rh$_2$O$_{10}$] units, B1 and B2/B3 (see text), are found experimentally
\cite{Li2RhO3_structure_2011}.
Results of spin-orbit MRCI calculations are shown, with a {\it local} coordinate frame for each
Rh-Rh link ($x$ along the Rh-Rh bond, $z$ perpendicular to the Rh$_2$O$_2$ plaquette). 
The form of the actual lattice spin model is detailed in the SM.
}
\label{Rh213_mgn_coupl}
\begin{tabular}{crr}
\hline
\hline\\[-0.25cm]
Energies \& effective couplings
&B1\footnotesize{$^1$}
&B2/B3\footnotesize{$^2$} \\
\hline\\[-0.15cm]
$E_2$ ($\Psi^{ij}_{\mathrm{2}}$)            &$0.0$     &$0.0$  \\
$E_3$ ($\Psi^{ij}_{\mathrm{3}}$)            &$2.5$     &$-3.3$ \\
$E_1$ ($\Psi^{ij}_{\mathrm{1}}$)            &$4.5$     &$4.6$  \\
$E_{\mathrm{S}}$ ($\Psi^{ij}_{\mathrm{S}}$) &$13.5$    &$1.9$  \\[0.20cm]
$J$                               &$-10.2$   &$2.4$  \\
$K$                               &$-2.9$    &$-11.7$\\
$\Gamma_{xy}$                     &$-1.3$    &$3.6$  \\
$\Gamma_{zx}\!=\!-\Gamma_{yz}$    &$ 2.8$    &$ 1.6$ \\
\hline
\hline
\end{tabular}\\
\footnotesize{$^1$ $\measuredangle$(Rh-O-Rh)=93.4$^{\circ}$, $d$(Rh-Rh)=2.95 ($\times$2), $d$(Rh-O$_{1,2}$)=2.03 \AA}.\\
\footnotesize{$^2$ $\measuredangle$(Rh-O-Rh)=94.1$^{\circ}$, $d$(Rh-Rh)=2.95 ($\times$4), $d$(Rh-O$_{1}$)=2.03, $d$(Rh-O$_{2}$)=2.00 \AA .
O$_1$ and O$_2$ are the two bridging O's.}
\end{table}

\begin{figure}[!ht]
    \hspace{1.5cm} \includegraphics[width=1.0\columnwidth]{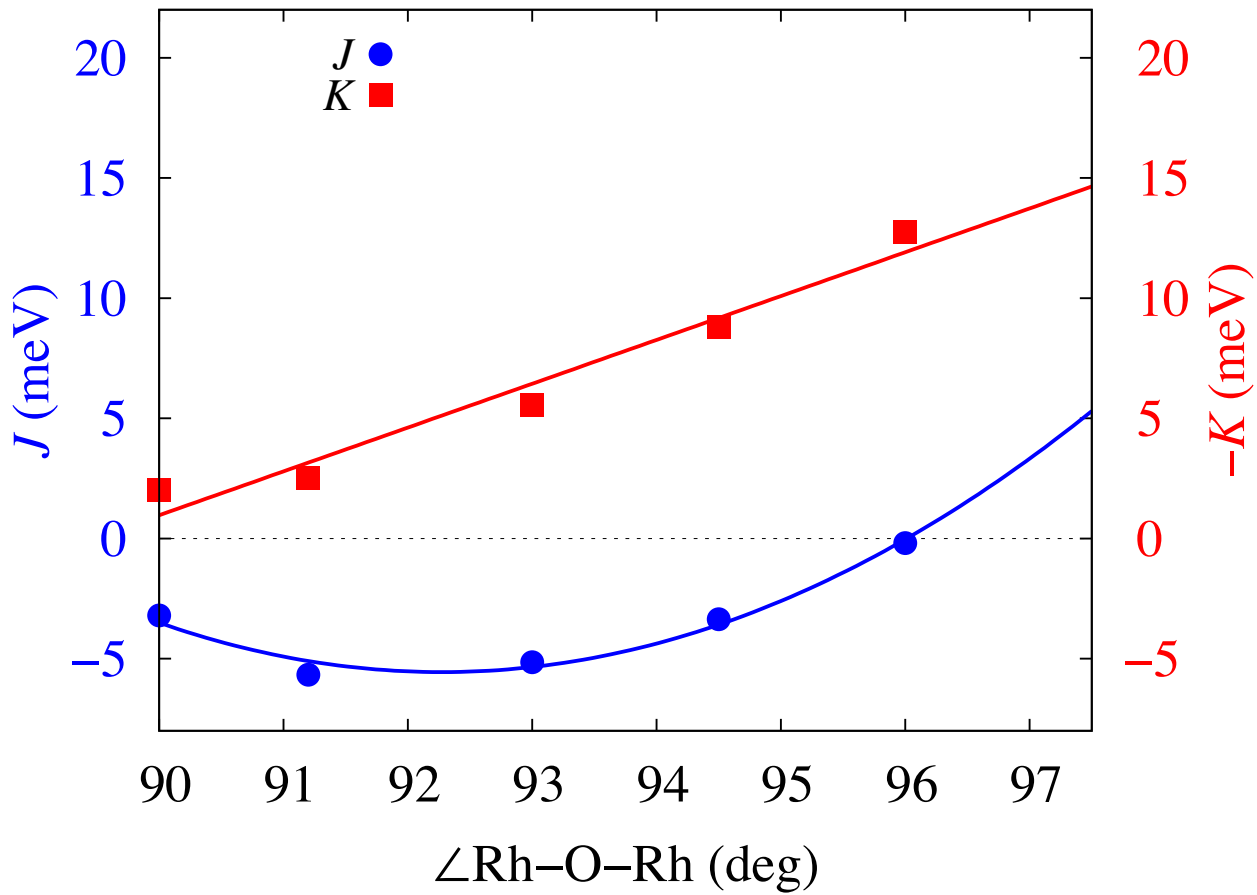}
\vspace{1.0cm}
\caption{
Dependence of the NN $J$ and $K$ on the Rh-O-Rh bond angle for an idealized structural model 
where all Rh-O bond lengths are set to the average value in the experimental crystal structure
\cite{Li2RhO3_structure_2011}.
MRCI+SOC results are shown.
The variation of the Rh-O-Rh angles is the result of gradual trigonal compression of the O
octahedra.
}
\label{idealstructures}
\end{figure}

\begin{figure}[!ht]
\includegraphics[width=1.0\columnwidth]{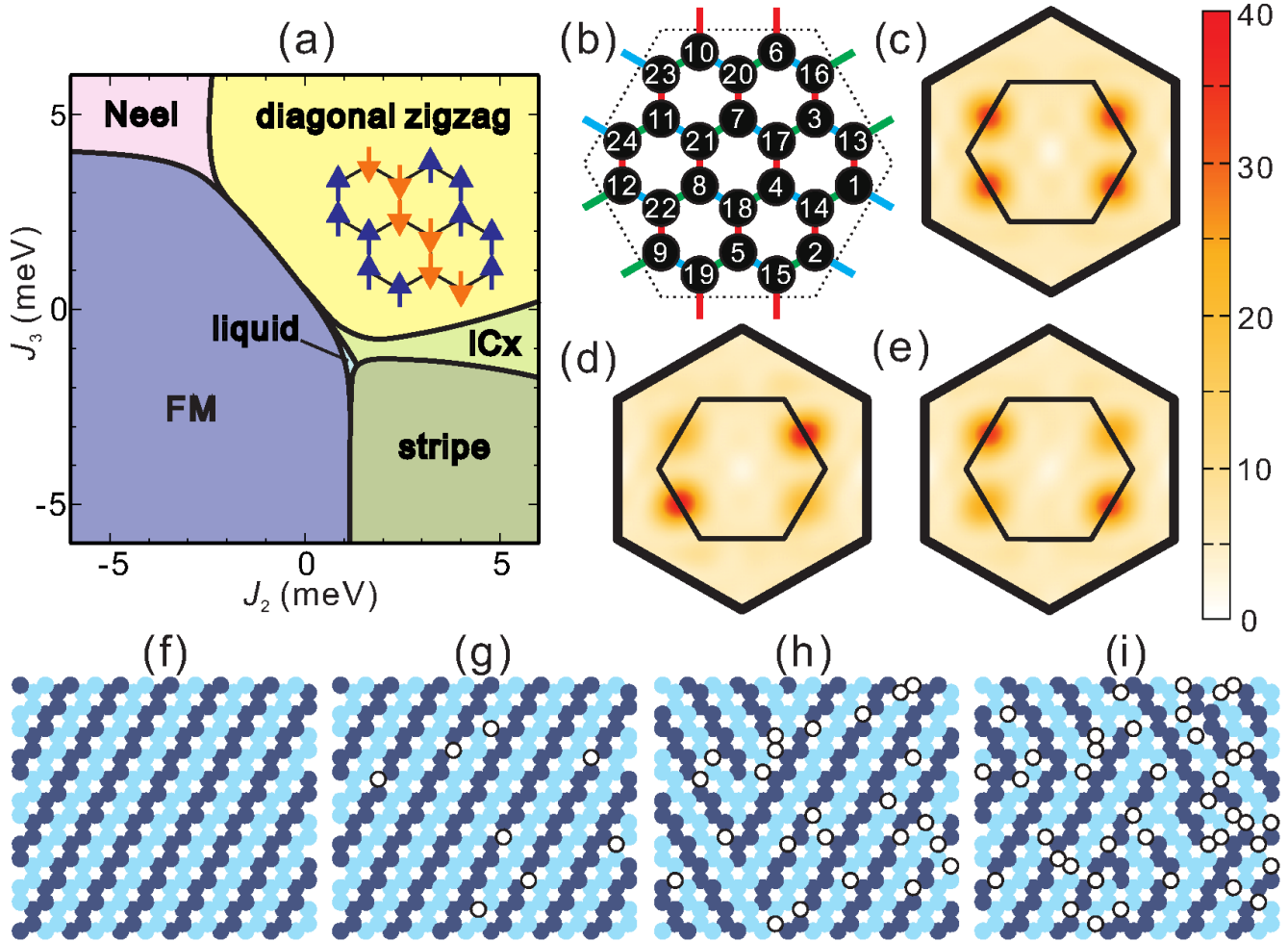}
\caption{
(color online)
(a) Phase diagram for the effective model of Eq.\,(1) supplemented by 2nd- and 3rd-neighbor
isotropic couplings $J_2$, $J_3$.
The NN effective interaction constants are set to the QC values provided in Table\,II.
The spin structure for the diagonal zigzag state is also shown.
(b) Sketch of the cluster used in the ED calculations;
the site index $p$ runs from 1 to 24.
Spin structure factors for $J_2\!=\!J_3\!=\!3$ with either
(c) no spin defects or 
two spin defects at
(d) $(p_1,p_2)\!=\!(17,20)$ and
(e) $(p_1,p_2)\!=\!(17,18)$.
Spin configurations for the simplified Ising model are shown for spin-defect concentrations of
(f) $x\!=\!0$,
(g) $x\!\sim\!2\%$,
(h) $x\!\sim\!7\%$, 
(i) $x\!\sim\!12\%$. 
Filled dark and light circles indicate opposite spin directions.
Open circles show the position of spin defects.
}
\label{phasediagram}
\end{figure}

\end{document}